%% file: main.tex
\documentclass{aastex7}

\newcommand{\um}{$\mu$m}
\newcommand{\spherex}{SPHEREx}

\accepted{July 5, 2026}
\submitjournal{ApJS}

\begin{document}

\title{The SPHEREx image and spectrophotometry processing pipeline}


\input{author_list_l1-3_pipeline_paper}

\begin{abstract}

In this paper, we describe the \spherex\ image and spectrophotometry data processing pipeline, an infrastructure and software system designed to produce calibrated spectral images and photometric measurements for NASA’s
\spherex\ mission.  \spherex\ is carrying out a series of four all-sky spectrophotometric surveys at 6$\farcs$15 spatial resolution in 102 spectral channels spanning 0.75 to 5 \um.   The pipeline 
which will deliver the flux- and wavelength-calibrated data products deriving from these surveys has been developed and is operated by the \spherex\ Science Data Center at Caltech/IPAC in collaboration with the \spherex\ Science Team.  Here we describe the framework and modules used in the pipeline, along with the data products, which are available at the NASA/IPAC Infrared Science Archive.

\end{abstract}


\section{SPHEREx Mission Overview} 

\spherex\footnote{\url{https://spherex.caltech.edu}}, the Spectro-Photometer for the History of the Universe, Epoch of Reionization, and Ices Explorer~\citep{Bock2026}, is a NASA medium explorer mission which launched on 11 March, 2025 and started science operations on 1 May, 2025. \spherex\ is conducting an all-sky near-infrared spectral survey in 102 spectral channels spanning 0.75 and 5 \um. During its 25 month prime mission, \spherex\ will survey the whole sky 4 times, and make hundreds of  observations per spectral channel in deep fields at the north and south ecliptic poles.  The \spherex\ Science Team will focus on three science cases: using galaxy redshifts to map the 3-dimensional structure of the universe to constrain models of inflation; using intensity mapping to trace the history of star formation in galaxies; and mapping the distribution of water and biogenic molecules in our galaxy \citep{Bock2026}.   The \spherex\ observations will result in a unique all-sky data set that will be available to the community via the NASA/IPAC Infrared Science Archive (IRSA)\footnote{\url{https://irsa.ipac.caltech.edu}}.  These data will enable the global astrophysics community to investigate science questions ranging from Solar System Objects to high-redshift galaxies \citep{Dore2016, Dore2018}. 

SPHEREx carries a single instrument designed to maximize spectral throughput and efficiency \citep{Korngut2026}. The focal plane is organized into two sides, both of which are configured with an arrangement of 3$\times$1 H2RG 
detector arrays.  Each detector has been overlaid with a linear variable filter, and is therefore sensitive to a specific wavelength band, and  
covers 3.5$\times$3.5 degrees on the sky with 6$\farcs$15 pixels.  A dichroic beam splitter separates the short-wavelength and long-wavelength bands:
the short-wavelength side (bands 1, 2 and~3) covers 0.75–2.44 $\mu$m, while the
long-wavelength side (bands 4, 5 and~6) covers 2.40–5.01 $\mu$m.  Bands 1 and 4, 2 and 5, 3 and 6 cover the same region of sky, so each observation produces two wavelengths for each point in the field of view.  Each detector pixel samples a single wavelength, see \citet{Hui2026} and \citet{Ashby2026} for more details on laboratory and on-sky calibrations.

The SPHEREx mission uses a custom Survey Planning Software (SPS, \citealt{Bryan2025}) to plan its observations.  The SPS is designed specifically to meet the mission’s unique requirements for both all-sky and deep field surveys. The SPS plans observations to maximize science time, to follow avoidance constraints imposed by the Sun, Moon, and Earth, and to maintain power through proper solar panel alignment. Observations are structured around “target groups,” which consist of multiple pointings that build complete spectral and spatial coverage across the sky, with overlapping fields to provide redundancy and mitigate errors. Each target group is separated by a large slew, while small slews separate observations within a target group.  The small slews are offset by 17' to observe consecutive spectral channels, resulting in efficient spectral sampling. The survey strategy is divided into four main surveys, defined by spacecraft orientation, and offset in later surveys to achieve Nyquist sampling of spectra.  The SPS is run for each planning period, generally twice per week.

The observations produced by SPHEREx are processed into science-ready, calibrated data products by the SPHEREx Science Data Center (SSDC), a hardware, software, operations and analysis system.  This paper describes that pipeline, focusing on the overall architecture and providing a high-level description of the individual algorithms, which are implemented in software as separate modules.  This is intended to document the fixed top-level properties of the pipeline and the relationship of its various components.  We have also authored an online SPHEREx Explanatory Supplement that is a living document updated and with a \added{change log} for each version of the pipeline used to generate public data products.  The Explanatory Supplement focuses on the implementation details of the individual modules, the provenance of the calibration products for each data release, and module validations.  The most recent version can be found at IRSA\footnote{\url{https://irsa.ipac.caltech.edu/data/SPHEREx/docs/overview_qr.html}}.  At the time of writing, the pipeline version is 6.4.

Section \ref{sec:pipe_overview} presents a high-level overview and describes the software framework used. Section \ref{sec:raw} describes the data ingestion process and Section \ref{sec:modules} describes the processing and data product generation.  Section \ref{sec:dataproducts} covers data products, including calibrations and Section \ref{sec:tools} describes SPHEREx-specific tools available at IRSA.  Section \ref{sec:operations}
covers the operations of the pipeline and  Section \ref{sec:future} includes some planned upgrades.

\section{Pipeline Overview}
\label{sec:pipe_overview}

The complete SPHEREx data processing path includes five distinct processing levels.  The high-level processing function of each level is:

\begin{itemize}
    \item {\bf Level 0:} Raw analog-to-digital outputs are processed within the instrument electronics on board, including computing slopes, detecting transients, and compressing and packetizing the data.
    \item {\bf Level 1:} Level~0 pixel slopes received on the ground are converted into engineering units, coarse astrometric solutions based on spacecraft pointing are computed, basic data quality checks are performed, and images are packaged into standard-format FITS files.
    \item {\bf Level 2:} Photometric and astrometric calibrations are calculated and applied to create calibrated and aligned spectral images and data cubes and data quality checks are performed.
    \item {\bf Level 3:} Forced photometry is performed at predefined reference positions, yielding wavelength-tagged photometric measurements, which are assembled into catalogs.
    \item {\bf Level 4:} High-level science data products are produced, including galaxy redshifts, ice column densities, and deep image mosaics.
\end{itemize}

The data levels are summarized in Table \ref{tab:levels}, while the products themselves are described in more detail in Sec. \ref{sec:dataproducts} and the Explanatory Supplement.  This paper focuses on data ingestion and Levels 1 to 3.

\begin{deluxetable}
{lll}
\tablecaption{SPHEREx Data Levels \label{tab:levels}}
\tablehead{
\colhead{Level} & \colhead{Format} & \colhead{Description}
}
\startdata
0 & Spacecraft packets & Instrument and spacecraft data \\
1 & FITS & Formatted photocurrent images in engineering units \\
2 & FITS & Calibrated spectral images \\
3 & Apache Parquet & Spectro-photometric measurements \\
4 & FITS and Parquet & Science catalogs and images \\ \hline 
\enddata
\end{deluxetable}

\subsection{Pipeline Flow}

\begin{figure*}[htb]
\centering
\includegraphics[width=5in]{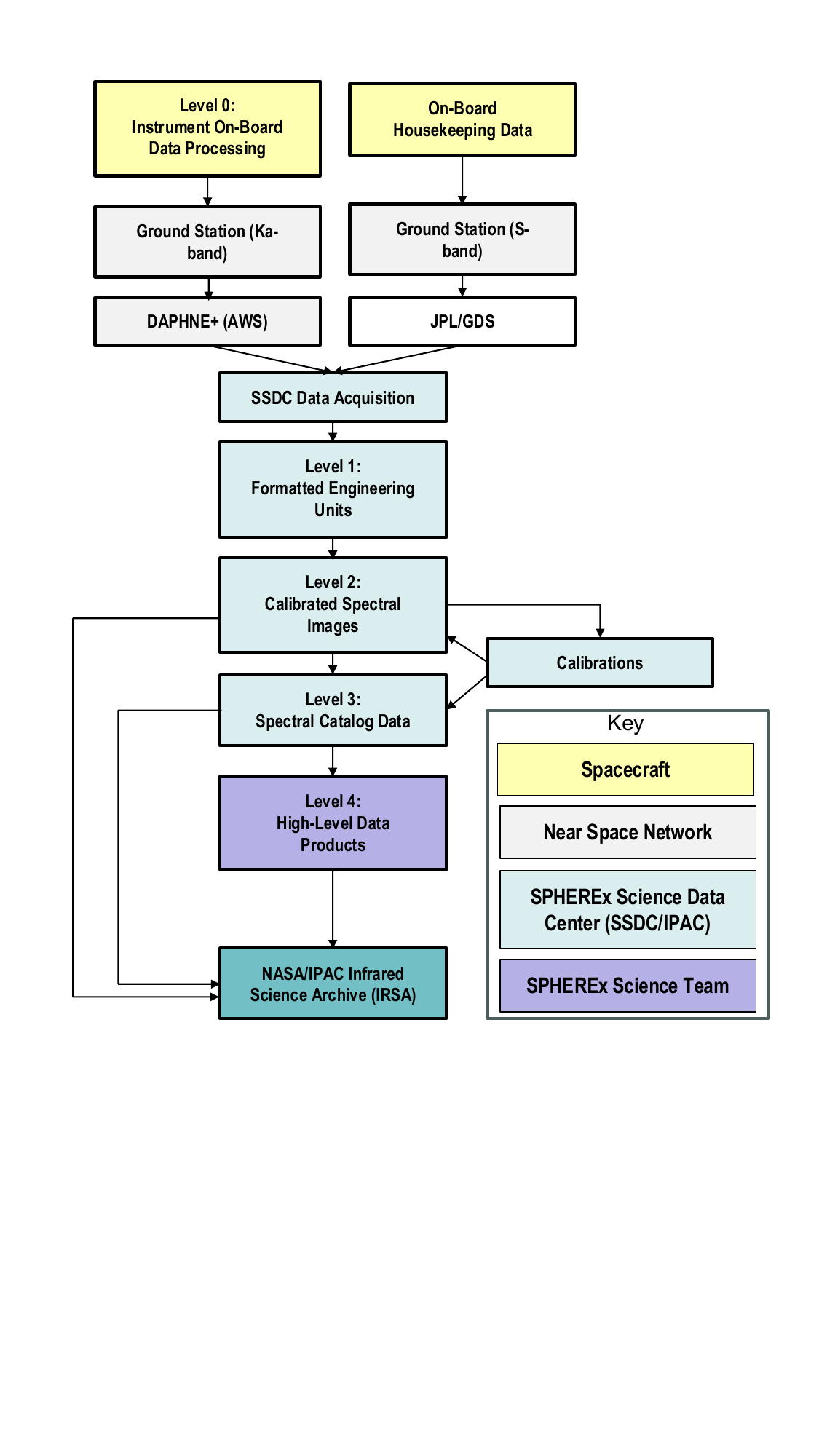}
\caption{The high-level flow diagram for the SPHEREx science processing pipeline
\label{fig:flowchart}}
\end{figure*}

As shown in Figure \ref{fig:flowchart}, several project elements are responsible for different aspects of the overall data flow and processing.
The definition and testing of the algorithms in Levels 1 to 3 are a collaboration between the SSDC and the SPHEREx Science Team, while the implementation and operations are the responsibility of the SSDC.  The SPHEREx Science Team is responsible for Level 4.  The Level 0 processing is performed by flight software onboard and was developed by the SPHEREx Instrument Team \citep{Korngut2026}.  Pre-launch, the pipeline was tested using simulated data products produced by the SPHEREx Sky Simulator \citep{Crill2025} and an additional simulator of the Level~0 processing, provided by the instrument team.  Post-launch, the pipeline was used during in-orbit checkout and in science operations.

The processing flow begins with the ingest of observation plans from the \spherex\ Ground Data System (GDS) at JPL, and of Level~0 data obtained from the spacecraft via two paths, a high-rate path for pixel data through the NASA Near Space Network (NSN) Data Acquisition Processing and Handling Network Environment (DAPHNE\verb|+|) system, and a medium-rate path for spacecraft and instrument housekeeping telemetry through legacy NSN interfaces and the GDS.
The ingest processing occurs in an automated, data-driven way triggered by the arrival of datasets, described in additional detail below.

Once the data have been ingested, processing proceeds with Level~1, handling images as they are fully assembled.  Level~1 executes independently on each image.

The next stage of processing, Level~2a, includes modeling of persistent-charge effects from image to image, and must therefore be run in order of exposure acquisition for each of the six bands.
Level~2b can then be run, without further ordering constraints, to create calibrated single-epoch images, and consumes the bulk of the pipeline computing resources up to this point.
Up through this \added{stage}, all the pipelines are run on-premises at IPAC, and all processing follows the availability of Level~0 data as closely as is practicable.
The Level~2b outputs are the quick release calibrated spectral images available at IRSA.

Forced photometry is performed on calibrated images in order to extract per-object spectra.
Level~3a performs the per-image processing to extract spectrophotometric measurements, and the final step, Level~3b, aggregates the resulting catalog-oriented data into per-object spectra.  

The Level 4 pipelines use products from both Level 2 (calibrated spectral images) and Level 3 (spectrophotometric measurements) and these Science Team-developed and operated pipelines will be described in future publications.

\subsection{Framework}

The Level 1 to 3 pipelines are implemented, primarily in Python, as an application of the open-source data access and pipeline construction middleware first developed for the Vera C. Rubin Observatory.  Here we describe the general structure of that middleware and how it was used for SPHEREx.

The Rubin data access framework (known as the ``Butler'') provides abstractions that isolate the scientific code in the pipelines from a number of common concrete details \citep{Jenness2022ButlerPipeline}.
The Butler handles the following tasks: interfacing with underlying storage systems (Posix and cloud storage), file and directory naming, serialization and de-serialization of Python objects, and the expression of relationships between datasets (e.g., the association of an image with calibration data appropriate for its processing).
This limits the pipeline module implementations to having to deal with the actual processing algorithms.
Serialization and de-serialization connecting physical datasets to Python objects is based on Butler `dataset types', discussed below. At the end of the ingest process all the resulting data are under Butler management.
Each dataset type is associated with a set of `data coordinates' that identify them within the universe of \spherex\ data.
Coordinates include items such as identifiers for pointings of the observatory, detector (band) numbers, and HEALPixel tiles on the sky.
The Butler maintains dataset metadata in a relational database (PostgreSQL for \spherex), paired with a bulk data store which for \spherex\ is currently Posix-based (though future work may also make use of cloud storage).

The Rubin pipeline construction framework \citep{Jenness2022ButlerPipeline} provides a uniform base-class interface, `\texttt{Task}', for pipeline modules, including support for run-time configuration parameters.
The algorithmic code in concrete subclasses of this interface operates only on Python objects and does not need to concern itself with input and output.
This greatly facilitates debugging and the construction of testing frameworks.
The framework further provides a uniform configuration, a `\texttt{PipelineTask}', for the declaration of the principal inputs, required calibrations and other auxiliary inputs, and outputs of each stage of a pipeline, and for the definition of pipelines composed of multiple stages.
Finally, the framework provides uniform tools for the execution of pipelines on specified inputs and/or for the production of specified outputs.
These tools support the creation of directed acyclic graphs for the execution of a pipeline on specified data, and their subsequent execution, and they collect provenance metadata for the outputs of processing.  Each of the pipeline modules described below are implemented as a subclass of \texttt{Task}, and the major segments of the pipelines (Level~1, Level~2a, Level~2b, Level~3a, and Level~3b) as subclasses of \texttt{PipelineTask}.

Application of these frameworks significantly facilitated the construction of the pipelines in a uniform and maintainable way.
The use of common interfaces throughout the pipelines has made it easier for the SSDC developers to review and support each other's work, and makes the pipelines usefully self-documenting.  The use of the Rubin middleware for SPHEREx involved the creation of a dedicated data model for dataset types and data coordinates, successfully accommodating concepts specific to the space observatory environment and the SPHEREx instrument, stretching the software into a substantially new domain.
The experience supports using the middleware for other future missions.

\section{Level~0 Data Handling}
\label{sec:raw}

The SPHEREx ingest system provides the bridging layer between spacecraft downlink products and the Level~1–3 pipelines. It creates planned-observation records, coordinates asynchronous Ka-band and S-band data deliveries, validates data completeness, manages permanent archiving and staging areas, assembles initial \texttt{raw} exposure FITS files, and associates these and ancillary data with the planned observations.

The \spherex\ project has adopted Apache Kafka \citep{Sax2018Kafka} as its platform for messaging between major components of the ground system, including for operations associated with data ingest and the return of data quality information to the GDS.
The \spherex\ GDS maintains a Kafka server at JPL and the project has established a dedicated secure tunnel between JPL and IPAC for communication with this server.

\subsection{Instrument Processing and Data Transmission}

Within the \spherex\ instrument, as part of on-board Level 0 processing, six readout boards (one per detector) read out photocurrents, compute pixel-level slopes from detector readouts, measure preamplifier drifts and detector offsets, detect glitches, compress and packetize the data using the Space Packet Protocol\footnote{ CCSDS 133.0-B-1, Space Packet Protocol,\url{https://ccsds.org/Pubs/133x0b1s.pdf}}.  These packets contain 1010 bytes and are transferred to the spacecraft for temporary storage.
The spacecraft then wraps these packets into Sync-Marked Transfer Frames\footnote{CCSDS 732.0-B-3, AOS Space Data Link Protocol,\url{https://ccsds.org/Pubs/732x0b3e1s.pdf}} of 1024 bytes each and telemeters these frames to ground stations, via a 600 Mbps Ka-band link.  This transfer occurs multiple times per day (typically 1–5) through the NASA NSN. The high-rate Ka-band stream is processed, stored, and distributed by NSN's DAPHNE\verb|+|, hosted in the NASA Mission Cloud Platform (MCP) on Amazon Web Services (AWS).

For each Ka-band downlink received on the ground, DAPHNE\verb|+| produces (1) a set of binary bulk-data files, (2) a List of Delivered Files (LDF) manifest enumerating filenames, sizes, and checksums, and (3) an Accounting Report (ARP) summarizing the completeness of received frames. 
These products are first deposited into a private NSN bucket on the MCP, and then replicated into a mission-accessible User Mission Ground System (UMGS) bucket. By design, the LDF is written only after all bulk data are present, providing an atomic indicator that the delivery of data from a downlink is complete.
The project has established a trigger on the UMGS cloud bucket, which sends a message via Kafka for the arrival of new LDF and ARP files, using the mission's Kafka server at JPL.
 
In parallel, instrument and spacecraft housekeeping and ancillary engineering information are collected by the spacecraft and, during ground station passes, downlinked via a 2 Mbps S-band data link.  The resulting data files are transferred to the GDS at JPL via a legacy NSN interface.
The telemetry-derived data products include several time-series data types: a subset of the original S-band telemetry channels, as needed to support the pipeline code and populate fields in the output file headers; reconstructed spacecraft orientation, position, and velocity data; and ground-track information.  Each of these data types has an associated Kafka ``topic'' (essentially, a named channel), and data are delivered at various cadences as the GDS processes the downlinks it receives.

\subsection{Data Ingestion at IPAC}

The SSDC initiates ingest upon receipt of a Kafka notification that a new LDF is available in the UMGS bucket. The Kafka consumer at SSDC retrieves the LDF, ARP, and all referenced binary data files, storing them under a permanent downlink identifier that includes the pass and antenna IDs. For each ingest event, SSDC generates a File Receipt Report (FRR), returns it to mission operations via a dedicated Kafka topic, and archives it alongside the downlink metadata.  Intermediate products (e.g., unpacked instrument blocks) are retained in short-lived staging areas, while the original as-received bulk data files and metadata are archived indefinitely.

In parallel, the GDS transmits data to the SSDC via Kafka, including semi-weekly Predicted Events Files (PEFs), and a set of derived data products from the S-band telemetry downlinks.  A PEF contains one planning period~--- 3.5 days~--- of time-stamped commands to be executed by the spacecraft, including the actual imaging commands produced from the output of the Survey Planning Software.
The SSDC extracts these commands, which include unique Observation IDs, times, pointing boresights, and a set of metadata flags, and creates a per-planned-observation record in the Butler for each one.  PEFs are generally sent to the SSDC about two days in advance of the start of the time period of on-orbit operations they cover.  The SSDC uses the planned-observation records derived from the PEF to divide the time series into per-observation snippets, which are stored in the Butler.

Due to limited uplink bandwidth, the Observation ID is not round-tripped through the spacecraft and therefore neither the received pixel data nor the S-band derived data are inherently tagged with this ID.
The received data must therefore be reassembled and labeled by the SSDC using the relationship between timestamps and Observation IDs derived from the PEF.

Both Ka-band staging and GDS processing are triggered automatically upon receipt of the associated Kafka message. 
Further processing of the Ka-band image data includes creating a per-downlink frame-index file, which organizes received frames into continuous chunks and records, for the first and last frame of each chunk, the Extended Virtual Frame Count (EVCFC), packet timestamp, ground-receipt time, and byte offset to the DAPHNE header.
The newly created frame index is then reconciled with the combined frame index from the previous downlink to eliminate duplicates and overlaps with earlier passes. 
Note that the 28-bit EVCFC, which increments continuously on board, is sufficiently wide that roll-over back to zero occurs after more than twice the number of frames that can be stored in the on-board ring buffer.
Therefore, every frame in memory at any given time has a unique EVCFC value; this is a valuable property in the deduplication analysis.
From the resulting deduplicated view (the updated combined frame index), the system produces, for each detector, a brief packet-level summary and a continuous stream of compressed science data for that pass. 

A high-water-mark check evaluates storage backlog against the known size of the on-board solid-state recorder to decide which missing data chunks might still be receivable and which must be expired --- treated as permanently lost.
Next, packet timelines are matched to the planned-observation records, and each segment is tagged with the correct Observation ID. The system then generates assembly instructions to build exposure-level data products for complete observations;  observations with missing data can be expired and released to processing if policy requires. Using these instructions, \texttt{raw} exposures and their associated image metadata are assembled and registered in the Butler data repository for Level~1 processing. 

Aside from the requirement that planned-observation records exist in advance (i.e., the PEF must be received before any of the corresponding observations occur), there is no constraint on the order in which data arrive at the SSDC, and duplicate data deliveries are handled transparently. Once both Ka-band and S-band ingest steps are complete for a given time period, generally one day of operations, SSDC launches the Level 1 pipeline.

\section{Pipeline Organization and Modules}
\label{sec:modules}

Levels 1 to 3 of the pipeline are organized into modules.  Each module has a specific function in the processing flow, has defined inputs and outputs, and is implemented as \texttt{Task} in the framework.
Details for each algorithm are available in the SPHEREx Explanatory Supplement, which is updated for each version of the pipeline used for public data products and 
\added{the python module code and test directories are available in github (https://github.com/Caltech-IPAC/spherex-pipelines) for the community to see the exact implementation.}   Here we summarize the module functionality in each level.  

\subsection{Level 1}
\label{sec:Level1}
The goal of the Level 1 pipeline is to take the Ka-band and S-band inputs from data ingestion and produce engineering unit images in a standard format and with the necessary header information to enable later calibrations to physical units.

The Level 1 pipeline is made up of the following seven modules:
\begin{itemize}
    \item GDS Image Association: Merge detector data from the Ka-band downlinks with header values from the S-band downlinks.
    \item Estimate Coarse Astrometry: Estimate initial pointing information based on the survey plan and geometric offsets between detectors, with an accuracy of $2^{\prime}-10^{\prime}$.
    \item Validate Sample-up-the-ramp (SUR): Extract the 8 $\times$ 34 pixel subsection of the detector where the full sample up the ramp is recorded to monitor noise performance and temporal stability. 
    \item Convert to Engineering Units: Convert Level~0 pixel slopes from analog digital units (ADU) to electrons per second using gains provided by the Instrument Team.
    \item Correct Preamplifier: Apply corrections for amplifier chain drifts using reference pixels and interleaved phantom pixels. The role of the reference pixels is described in \cite{Nguyen2025}.
    \item Create Level 1 Image Data Files:  Assemble a FITS file with the engineering unit detector data and appropriate header information.
    \item Data Quality Assessment for Level 1: Perform basic checks on the data integrity, including gathering image statistics.  Results from this assessment are recorded as header keywords.
\end{itemize}

\subsection{Level 2}
\label{sec:level2}
The Level 2 processing includes astrometric and photometric calibrations using a combination of ancillary catalogs, internally generated calibration products from the pipeline, and calibration products from ground testing. The Level~2 processing is run in two segments, Level 2a, which must be run in the order of image acquisition for each detector, as it includes taking into account the effects of persistent charge retention from one image to the next, and Level 2b, where the end product is calibrated spectral images to be used for science.

The Level 2a pipeline consists of these three modules:
\begin{itemize}
    \item Correct Non-linearity: Corrects for detector non-linearity, which arises from declining gain as charge accumulates within each H2RG pixel \citep{Zengilowski2020}.  This is applied to all pixels except for reference and phantom pixels.
    \item Create Variance Map: Generate a per-pixel variance map using the flux value from Level 1, using the method described in \cite{Robberto2007}.  The variance is an image extension in the Level 2 spectral image (Sec. \ref{sec:dataproducts}).
    \item Update Persistence: Apply a parameterized analytic model \citep{Fazar2025, Fazar2026} to estimate the level of persistent flux in each pixel based on the cumulative flux exposure of that pixel, and flag pixels significantly affected by persistence from earlier images.  
\end{itemize}

Following the completion of Level~2a, the remainder of the pipeline processing can be performed on images in any order and at any level of parallelism.

After Level 2a, the reference and phantom pixels are removed from the FITS files as they are not needed for subsequent processing steps and do not reflect photons from the sky.  At this point in the processing, the pipeline system also executes the Query Reference Catalog module which identifies the subset of sources from the SPHEREx Reference Catalog that should be covered by an image's pointing, and evolves objects with known astrometric motions to the epoch of the observation.
The SPHEREX Reference Catalog is assembled periodically by the Science Team and includes photometric and astrometric calibrators, along with sources from the science themes for which forced photometry is to be performed in Level~3 \citep{Yang2025}.  To improve pipeline efficiency, the Query Reference Catalog module executes its query over the sky area covered by the overlapping images in the target groups mentioned above, rather than separately for each image.

The Level 2b pipeline is composed of the following nine modules:
\begin{itemize}
    \item Apply Bad Pixel Mask: Generate a per pixel map with a multi-bit flag tracking on-board information (e.g., transients, nonfunctional pixels, persistence and outlier pixels).  The complete list of flags and their assigned bit mask is given in section 3.2.4 of the Explanatory Supplement. The flags are an image extension in the Level 2 spectral image (Sec. \ref{sec:dataproducts}). 
    
    \item Apply Dark, Flat Field and Photometric Calibration: Apply the dark current, flat field and absolute gain calibrations to the flux.  These calibrations are derived in separate modules (Sec. \ref{sec:cals}).  After this step data values in the images are in units of MJy/sr.
    
    \item Estimate Fine Astrometry: Calculate an astrometric solution to determine the translation and rotation of the field per image.  Astrometric calibrators are selected to have a good Gaia astrometric solution, SNR $>10$ but not predicted to be saturated in any SPHEREx band, and contain no sources within 2 SPHEREx pixels that are brighter than 10\% of the flux of the calibrator. 
    Two stages are used for the calculation, first using the \texttt{Astrometry.net}\footnote{\url{https://www.astrometry.net}} package \citep{langAstrometry2010} to derive an initial solution and then the \texttt{scamp} software \citep{Bertin2006}, which uses a similar approach as the WISE astrometry pipeline\footnote{\url{https://wise2.ipac.caltech.edu/docs/doc_tree/sds/sds-PRex.pdf}}, to obtain the final refined solution. Note that the latter is run iteratively, which leads to improved solutions.  The outputs are recorded as header keywords and a header keyword (\texttt{FINAST}) is added indicating astrometric performance, with a value of 0 for images with rms $< 1\arcsec$. The median astrometric alignment precision of images with a \texttt{FINAST} flag equal to zero reaches $0.1-0.4^{\prime\prime}$ on average.

    \item Compute Known Solar System Object Positions: Using S-band telemetry, WCS pointing information and the orbit propagation tool \texttt{kete} \citep{kete}, create a  list of Solar System Objects (SSOs) from the SSO catalog that fall in the current SPHEREx frame, their astrometric coordinates during the integration, and the UTC time of the observation. It may also include predicted fluxes for the sources to inform priors for forced photometry. This source list is stored in an ancillary parquet file.
    
    \item Create Source Mask:  Add a flag for pixels predicted to be impacted by flux from known sources.  The input source list is generated by the Query Reference Catalog module and the flag is added to the flag image extension populated by the Apply Bad Pixel Mask module.
    
    \item Detect Outlier Pixels: Detect and flag transient events that were not flagged by onboard procedures, using a sigma clipping algorithm applied to a median-filtered images.  The flag is added to the flag image extension populated by the Apply Bad Pixel Mask module.
    
    \item Get Zodi Model: Provide an estimate of the diffuse zodiacal light based on the spatial location of the image.  The empirical time-variant model is based on \cite{Kelsall1998}, updated with spectral measurements from \cite{Tsumura2013}. For a more detailed description of the model, see also \cite{Crill2025}.  The zodiacal model is an image extension in the Level 2 spectral image (Sec. \ref{sec:dataproducts}) and is not subtracted from the image but is provided for later operations, such as mosaicing.
    
    \item Create Spectral Images:  Create Spectral Image file in FITS format.  This format, which is unique to SPHEREx, is described in Section \ref{sec:dataproducts} and in the SPHEREx Explanatory Supplement.  The SPHEREx spectral images are the quick release and reprocessing image format available via IRSA.
    
    \item Data Quality Assessment for Level 2:  Perform a series of automated tests on the output from each module and on the final spectral image FITS file.  To be transferred to IRSA, the image must pass at least half of the module tests, be outside the South Atlantic Anomaly, have a transient count less than 435,000 pixels and have a fine astrometry flag value of 0, indicating a solution meeting requirements.
\end{itemize}

\subsection{Level 3}

The Level 3 pipeline is run as a scatter-gather operation in two phases.
Level~3a estimates backgrounds and performs forced photometry on each image, producing an output file for each Level~6 HEALPixel ($\texttt{NSIDE}=64$, approximately 1 sq. deg.) covered by the image.
Level~3b, which must be run after the completion of Level~3a for all data on a given region of sky, collects all the photometry results for a single HEALPixel tile, and consolidates them into the desired catalog format.

Level~3a consists of these modules:
\begin{itemize}
    \item Estimate Diffuse Background: Estimate the background level from source-masked L2 images.   Use either a smoothly-varying global background estimate derived by interpolating small sub-regions of the image, or a local background estimation.  The global background calculation starts with a spectral fit to the zodical light contribution in the image, and the residual is fit with the \texttt{Background2D} class from \texttt{Photutils} \citep{Bradley2024}. The combination of the spectral fit and the Background2D result is the global background estimate. This method is used whenever the image contains $>90$\% usable subregions. Subregions are rejected if $>$75\% of pixels in the regions are identified as containing non-negligible source flux from sources in the SPHEREx Reference Catalog.
    Local background estimation is used whenever the global estimate fails because of an unusably high source density. Local subtraction also uses \texttt{Background2D}, but applied to a 15 $\times$ 15 pixel region around a cutout. Local subtraction is typically triggered in the galactic plane for latitudes $\lesssim 20^{\circ}$.
    
    \item Perform Forced Photometry:  Extract SPHEREx photometry based on prior knowledge of source positions and shapes from the SPHEREx Reference Catalog \citep{Yang2025} using an algorithm based on the community package \textsc{Tractor} \citep{tractor16,weaver22}.  The module works by finding the maximum likelihood solution for source fluxes by comparing the measured image with a parametric model that predicts the scene given the calculated PSF and known source positions.  An improved error estimation procedure that accounts for covariance between blended sources was developed and implemented (see \citet{Huai2025} for details).
    This module first removes the background computed in the previous module.  This module operates on a 15 $\times$ 15 pixel cutout and generates a model containing all sources within that region.
    The fit flux is converted to $\mu$Jy using the Solid Angle Pixel Map (Sec. \ref{sec:calibrations}).
    As data quality indicators, flags from the pixels in the central source region are aggregated from the image and included in the output along with a fit metric.  More details on this metric are provided in the Explanatory Supplement.
\end{itemize}

Level~3b consists of the following modules, some or all of which may be run depending on the specific data processing task being performed:

\begin{itemize}
    \item Create Primary All-sky Catalog: Merge flux measurements from all L2 images into a single unified parquet catalog. For each source that is photometered on an L2 image, the All-Sky Spectral Catalog includes the measured flux, associated error, wavelength and time in modified Julian date of the observation. This product is used by the SPHEREx Science Team as input to Level 4 products and as input to the High Reliability Source Catalog.  
    
    \item Create Secondary All-sky Catalog and Estimate Variability: Create a secondary all-sky catalog from the primary all-sky catalog with the spectrum from each source averaged over the observations from all surveys to date, and interpolated to a fixed wavelength grid.  Estimate which measured sources from the SPHEREx Reference Catalog are variable. Two methods will be used, depending on the S/N of the source: in both cases, compare the current photometric measurements to previous SPHEREx measurements of the same source and apply a specified sigma noise-threshold. For bright (S/N $>$ 100) sources, perform the comparison in the native spectral channels, while for faint sources, collapse the spectrum to a limited ($\sim$10) number of reference bands.  This product requires at least two surveys of input data to provide valid wavelength gridding and provides input to the High Reliability Source Catalog. 
    
    \item Create High Reliability Source Catalog: Select a subset of the all-sky catalog based on SNR and survey to survey consistency (see Sec. \ref{sec:dataproducts} for a description of the catalog).

\end{itemize}

\subsection{Calibrations}
\label{sec:calibrations}
The calibrations applied in the pipeline modules summarized above are calculated in dedicated modules.  The frequency with which these modules are run varies with the expected variability and the input data required.
\begin{itemize}
    \item Measure Exposure-Averaged Point Spread Function:  Calculate the average point spread function (PSF) on an 11 $\times$ 11 grid averaged across hundreds of images for a given band. The 121 zones are chosen to track the spatial and wavelength-dependent variations of the PSF across a given band. The PSFs are constructed by stacking cutouts of individual stars in each of the zones. Several hundred exposures are used to obtain $>$1000 stars per zone and per band. The stacking is done at $10\times$ oversampled pixel scale as the PSF is significantly undersampled at SPHEREx resolution. In a last step, the resulting super-resolution stack is iteratively deconvolved by the pixel grid function, which takes into account the spread of the flux over
    the super-resolved pixels.
    
    \item Estimate Dark Current and Flat Field: Fit the slope of each pixel value in 
    electrons sec$^{-1}$ against a reference flux level measured from the median brightness in
    the pixel’s spectral channel in each image. This procedure uses 510 spectral channels per band, such that the flux of the background in each channel is expected not to significantly vary in the
    spectral dimension.   An additional tilt to the flux in each spectral channel is applied to account for solar elongation, i.e., enhanced
    brightness of the zodical light background on one side of the band.
    The slope of the relation over
    many images for each pixel represents the relative pixel response, or flat field, within the spectral channel,
    while the y-intercept of the line fit gives the dark current in electrons sec$^{-1}$.  Roughly 1000 images are sufficient
    to yield a consistent and repeatable  dark current estimate and flat field response.
    
    \item Estimate Absolute Gain Matrix: Calculate a 1D gain
    function, i.e., the gain as a function of the spectral dimension on each detector using the measured vs expected flux for a set of pre-identified primary calibrator stars.  These stars are well-studied community standards and are listed in the Explanatory Supplement.  Individual spectral measurements are convolved with the known bandpass and color-corrected.  Many thousands of measurements are combined using statistical weights to produce the gain function. This gain function is then
    multiplied by the inverse of the derived flat-field response, yielding the absolute gain matrix.
\end{itemize}
\label{sec:cals}

\section{Data Products}
\label{sec:dataproducts}

This section summarizes the data products, both public and internal, produced by the Level 1 to 3 pipeline.   Table \ref{tab:products} summarizes the public data products, sizes and availability at IRSA.   The data volume for each spectral image is 72 MB.  Descriptions of these products are given in Sec. \ref{sec:datadescr} and the production generation cadence is given in Sec. \ref{sec:cadence}.

\begin{deluxetable}
{lrl}
\tablecaption{SPHEREx Public Data Products \label{tab:products}}
\tablehead{
\colhead{Product} & \colhead{Total Data Volume} & \colhead{Availability at IRSA} \\
}
\startdata
Spectral Images & 190 TB & Within 60 days of data acquisition\\
All-sky Data Cubes & 95 TB & Year 1 release (Nov 2026); Year 2 release (Nov 2027) \\
High Reliability Source Catalog & 120 TB & July 2027 \\ \hline \hline
\multicolumn{3}{c}{Calibration Products} \\
\multicolumn{1}{c}{Product} & Per File Size & \multicolumn{1}{c}{Input Source} \\ \hline
Absolute Gain Matrix & 17 MB & Spectral Images \\
Exposure-Averaged PSF & 5 MB & Spectral Images \\
Dark Current & 17 MB & Spectral Images \\
Spectral WCS & 33 MB & Laboratory Data \\
Dichroic Impacted Pixels & 4 MB & Laboratory Data \\
Nonfunctional Pixels & 34 MB & Laboratory Data and Spectral Images \\
Non-Linearity Parameters &  83 MB & Laboratory Data \\
Electronic Gain Factors & 5 KB & Laboratory Data \\
Read Noise Parameters & 50 MB & Laboratory Data \\
Solid Angle Pixel Map & 16 MB & Laboratory Data \\
\enddata
\tablecomments{All calibration file types are available at IRSA and are updated when new versions are used in the pipeline.}

\end{deluxetable}

\subsection{Data Product Descriptions}
\label{sec:datadescr}

Here we briefly describe the currently produced data products along with those planned for later releases.  Table \ref{tab:products} lists those that are or will be available at IRSA.
Full descriptions of the spectral image file format are available in the Explanatory Supplement (Section 2, with a full listing of the header in Appendix A). 

\paragraph{Spectral Images}  These FITS files are the primary output of the Level 2b pipeline and are ready for use in scientific investigations.  As described in Section \ref{sec:level2}, these images have been astrometrically and photometrically calibrated and are in units of MJy\,sr$^{-1}$.  The pipeline generates one FITS file per band per exposure with the following extensions: \texttt{IMAGE}, \texttt{FLAGS}, \texttt{VARIANCE}, \texttt{ZODI}, \texttt{PSF} and \texttt{WCS-WAVE}.  At IRSA, users can search these files by spatial location, wavelength band, and time.  They can also be downloaded in bulk from IRSA on premises or from the cloud.  

\paragraph{Calibration Products}  The calibration products derived by the pipeline,  plus those supplied by the SPHEREx team and used by the pipeline, are available at IRSA.  These calibrations have been applied in the processing described above; the files are provided as a reference.
The calibration file version used to produce a given image file is recorded in the spectral image header.  Further descriptions of the calibration products are available in Section 4 of the Explanatory Supplement.

\begin{itemize}

\item{Absolute Gain Matrix:}
This product is one of the outputs of the Estimate Absolute Gain Matrix module (Sec. \ref{sec:calibrations}).  The product format is one FITS image file per band. 

\item{Exposure-Averaged Point Spread Function:}  The average PSFs calculated by the Measure Exposure-Averaged PSF module (Sec.~\ref{sec:calibrations}) are available both in this calibration file and as a layer in the spectral images.  The product format is one FITS file per band with dimensions $101\times101\times121$. 

\item{Dark Current:}  This product is one of the outputs of the Estimate Absolute Gain Matrix module (Sec. \ref{sec:calibrations}).  The product format is one FITS image file per band.

\item{Spectral WCS:} These files contain the central wavelength and bandwidth for each pixel. The central wavelengths are set to the point where the cumulative transmission curve reaches 50\% of its total area (i.e., the transmission curve median value for each pixel).
The bandwidths are computed as the central wavelength divided by the detector's measured spectral resolution. 
The product format is one FITS image file per band. The \texttt{CWAVE} and \texttt{CBAND} extensions contain per-pixel values based on laboratory measurements by the instrument team \citep{Hui2026} and should be used for science investigations.  The \texttt{WCS-WAVE} extension is a compact piecewise-bilinear approximation provided for visualization, following the FITS standard's \texttt{WAVE-TAB} WCS specification.

\item {Dichroic Impacted Pixels:}
This bit mask indicates pixels where the flux attenuation due to the dichroic filter is 50\% or higher and corresponds to the \texttt{DICHROIC} flag in the FLAG extension.  The product format is one FITS image file per band. 

\item{Nonfunctional Pixels:}
This bit mask indicates pixels known to be permanently non-functioning and corresponds to the \texttt{NONFUNC} flag in the spectral images' \texttt{FLAGS} extension. The pixel value is 0 for unaffected pixels and 1 for impacted pixels.  The product format is one FITS image file per band. 

\item{Non-linearity Parameters:}    
This file contains the input non-linearity calibration parameters used in the Correct Non-linearity module (Sec.\ref{sec:level2}).  The product format is one FITS image file per band. 

\item{Electronic Gain Factors:} 
These files include the per amplifier channel values used in the Convert to Engineering Units module (Sec. \ref{sec:Level1}). The product format is a single YAML (a human-readable data serialization language) file, which includes the provenance information for the detectors and a list of 32 gain values per detector. 

\item{Read Noise Parameters:} 
This file contains the per-pixel electronic noise used in the Create Variance Map module (Sec. \ref{sec:level2}). The per-pixel values are given in units of electrons and the product format is one FITS image file per band. 

\item{Solid Angle Pixel Map:} 
This product contains a measure of the solid angle per pixel in units of squared arcsec, which is used by the Perform Forced Photometry Module in the conversion from flux density to flux. The product format is one FITS image file per band. 

\end{itemize}

\paragraph {All-sky Data Cubes}  The all-sky spectral cubes will consist of full sky mosaics of SPHEREx image data, one for each of the 102 spectral channels, obtained by slicing the Spectral Image data into strips per spectral channel and recombining them into a mosaic. The spectral cubes will include the interpolated flux, variance, zodiacal light or background subtraction, exposure time/coverage, and bad-pixel mask flags.  These cubes will be produced as part of the Year 1 and Year 2 data releases.

\paragraph{High Reliability Source Catalog (HRSC)} The HRSC contains photometric measurements of highly statistically significant sources (SNR $>$ 10 in at least 5 of the 102 SPHEREx spectral channels) from the SPHEREx Reference Catalog that pass a survey-to-survey consistency test after the first year and a half of SPHEREx observations. This catalog includes both the primary and secondary measurement formats for each selected source as taken from the data generated by the Create Primary All-Sky Catalog and Create Secondary All-sky Catalog and Estimate Variability modules.

\subsection{Product Generation Cadence}
\label{sec:cadence}

The Level 2 calibrated spectral images are produced within a few days of data acquisition and are available at IRSA within 60 days of observation.  Following the completion of Year 1 (includes survey 1 and survey 2) and Year 2 (includes survey 1 to survey 4), all SPHEREx science data will undergo a complete reprocessing using the most up-to-date version of the Level 1-3 pipeline. These reprocessings, \added{called DR1 and DR2}, will incorporate the best available calibrations, processing parameters, and updated algorithms based on analysis of the initial processing results. 
Both spectral images and all-sky spectral cubes will be produced.
The HRSC will be produced after 1.5 years of survey data have been collected, after achieving Nyquist sampling, and updated as part of the Year 2 reprocessing.  The Year 1 and Year 2 products will be released at IRSA within 6 months of data acquisition and the HRSC within 8 months.

In addition to the above processing schedule, an additional reprocessing was performed in August and September 2025, based on considerable improvements in the PSF and the absolute gain over the initial data release in July 2025.  These data were released to the community via IRSA in October 2025 and are labeled QR2.  Details of the changes between the first data release (QR1) and QR2 are given in the SPHEREx Explanatory Supplement and we strongly recommend use of QR2 data for all science applications.  The Digital Object Identifier (DOI) for the QR1 images is 10.26131/IRSA629\footnote{\url{https://doi.org/10.26131/IRSA629}} \citep{irsa629} and for the QR2 images is  10.26131/IRSA652\footnote{\url{https://doi.org/10.26131/IRSA652}} \citep{irsa652}.

\section{Archival Tools}
\label{sec:tools}
In addition to the standard IRSA capabilities including data search, retrieval and visualization, IRSA will host several SPHEREx-specific tools.  Currently, the Spectrophotometry and Spectral Image Cutout tools are available, while the others are under development.  The Source Discovery tool will be available as a Python notebook and the others are available at IRSA via GUIs or application program interfaces (APIs). See the SPHEREx pages at IRSA for more details\footnote{\url{https://irsa.ipac.caltech.edu/data/SPHEREx/docs/overview_qr.html}}.
\begin{itemize}
    \item Spectrophotometry:  Measure spectra at user-supplied positions, using the same PSF photometry method based on \textsc{Tractor} as in the Level 3 pipeline but with user-supplied source positions instead of input from the SPHEREx Reference Catalog.
    \item Spectral Image Cutout:  Select sections of spectral image files based on user criteria (spatial area, band, time) and return a collection of images in the native pixel sampling.
    \item Source Discovery: Identify significant signal in user-supplied spatial region (with user constraints), without priors on the position.  
    \item Custom Mosaic: Create single-wavelength image(s) from spectral images using user supplied criteria, including synthetic bands.
    \item Spectral Cube Cutout:  Extract a subset from the All-sky Spectral Cube, returning FITS cube with HEALPix projection and optionally interpolate for synthetic bands.
\end{itemize}

\section{Operations}
\label{sec:operations}

Data ingestion as well as Level 1 and 2 processing tasks are run on-premises, using equipment at IPAC in the NASA Moderate Data Center (NMDC).  Level 3 processing is run at the Texas Advanced Computing Center (TACC).

\paragraph{Data Ingestion} GDS (S-Band telemetry) products are received via Kafka in the NMDC, and automatically processed in response to the Kafka message.  Similarly, Ka-band Level~0 pixel data ingestion from DAPHNE+ proceeds automatically in response to the receipt of the LDF-arrival message via Kafka, including depacketization and decompression and the creation of FITS image files.  Outputs from this processing in the Butler include planned observation records, per-observation GDS data records, and \texttt{raw} FITS files, and are tracked by SPS planning period.

\paragraph{Level 1} Beginning with the outputs of data ingestion, pipeline processing is steered by driver scripts using the Slurm workload manager\footnote{\url{https://github.com/SchedMD/slurm}} and processed in parallel across detectors.
The Observation Data Integrity Report (ODIR) aggregates the re-observation statistics, and is created after the L1 pipeline is complete. Currently the only trigger for a re-observation request is a transient flag count $>435,000$ in an image, which was set to produce roughly 1\% of images with re-observation requests. The list of these images is supplied to the GDS and the SPS attempts to re-observe them if possible.

The Query Reference Catalog module utilizes Butler dataset types: \texttt{level1} for Level 1 FITS images and \texttt{refcat} for the SPHEREx Reference Catalog. After performing its query, it generates a new dataset type named \texttt{srccat}, which contains all the sources for each dichroic pair and slew group, stored in Parquet files.

\paragraph{Level 2} The inputs to the Level 2 pipeline are \texttt{level1} FITS images and \texttt{srccat} outputs from Query Reference Catalog, as well as the needed calibration inputs. Level 2 Modules up to and including Update Persistence are executed sequentially by-observation in the NMDC, producing temporary \texttt{level2a} FITS files.
All remaining Level 2 modules are processed via Slurm jobs for maximum node usage and job management, and are additionally parallelized across detectors, producing \texttt{level2} FITS files. The Image Data Quality Report (IDQR) is created at the end of the completed L2 pipeline, includes information on per image pointing and PSF size.  This file is also returned to GDS via Kafka. Data quality information is saved in JSON files and may be used for trending analysis and cross checks.

\paragraph{Level 3} The inputs to the level 3 pipeline are \texttt{level2} FITS images, \texttt{srccat} parquet tables, and per-detector distortion map calibration FITS images. Since the time to photometer an image is proportional to the source density, the L3 execution time can vary greatly across the sky. To optimize TACC compute node usage, we create Butler \texttt{TAGGED} collections in the NMDC to group the \texttt{srccat} inputs based on the estimated Slurm job time. To support working without a database on the TACC machines, we generate quantum graphs, directed acyclic graphs which outline the corresponding \texttt{level2} and distortion map inputs for each \texttt{TAGGED} \texttt{srccat} collections. After transferring the input data and auxiliary files to TACC, we create serial Slurm jobs for each quantum graph. To photometer multiple images in parallel during a job, we rely on the LSST pipeline execution framework. Once the per-exposure, per-HEALPix output photometry measurement catalogs are generated at TACC, we transfer them back to the NMDC to combine source measurements across all detectors.

\section{Future Development Plans}
\label{sec:future}

As the SPHEREx team continues to work with and better understand the instrument and data, development work on the Level 1 to 3 pipeline will continue. 
In addition to the data products listed above, other areas which are currently under active development or discussion include:

\begin{itemize}
\item{Fine Astrometry:} Investigate if the fine astrometry performance is improved by calculating a fixed distortion from many images and calculating the differential changes for each image.  Investigate the ability to recover images with bad astrometry solutions by using the astrometric solution from other images offset by a small slew or other images within the same observation.

\item{Point Spread Function:} Investigate improvements on the measurement of the per-exposure PSFs. This may include a deconvolution by a kernel incorporating the settling jitter of the spacecraft (thought to be on the order of $\sigma_{\rm jitter}<0.1\arcsec$) or more accurate measurements by applying machine learning methods, e.g. \citet{Herbel2018}.

\item{Inter pixel response:}  H2RG detectors are known to have intra-pixel variations in detector sensitivity, e.g. \citet{Graet2022}.   This can be calculated by comparing the measured flux per pixel with the sum of the predicted contributions of individual sub-pixels and their responses but requires averaging over thousands of pixels across each detector.

\item{Improved Transient Flagging}:  Improve the flagging for transients, including earth-orbiting satellites and extended cosmic rays.

\item{Optical and Electrical Masking}: Improve the masking for a variety of optical and readout electronic effects known to be present in the SPHEREx Level-2 images, including: frame edge ghosts, due to light scattering from the optical filter mechanical structures; electrical crosstalk arising in the detector substrate and multiplexer; optical ghosts from reflections at the band defining beam splitter; other low-level optical ghosts from bright sources outside the field of view; and glow arising from large-angle response to the moon. Some of these effects were anticipated and characterized in the lab before flight and others were found during flight operations. 

\item Perform Data Quality Assessment for Level 3: Perform series of automated checks on the L3 data.

\end{itemize}

\section{Summary}
\label{sec:summary}

The SPHEREx Level 1 to 3 pipeline processes Level~0 spacecraft data through a structured, multi-level pipeline to transform onboard measurements into science-ready data products. The pipeline converts raw telemetry into engineering units, applies essential calibrations (astrometry, photometry, instrument corrections), and generates calibrated spectral images and catalogs of photometric, wavelength-tagged measurements. 

The system is built on modern, portable software frameworks, enabling trackable processing on both local and remote computing resources. Data are ingested daily from NASA’s ground networks and processed.  Science-ready calibrated spectral images are available to the community via IRSA, which also provides user tools to search, visualize and download SPHEREx data, as well as SPHEREx-specific tools for spectro-photmetric measurements, image cutouts, mosaics, and source discovery.

Operations are divided between IPAC’s data center (Levels 1–2) and the Texas Advanced Computing Center (Level~3), leveraging high-performance computing for large-scale photometric analysis. Future work will refine calibrations, expand science data products, and enhance community tools, ensuring continued availability of the unique SPHEREx datasets to maximize their scientific return.
 
\begin{acknowledgments}

We acknowledge support from the SPHEREx project under a contract from the NASA/Goddard Space Flight Center to the California Institute of Technology.
This research was partly carried out at the California Institute of Technology, under a contract with the National Aeronautics and Space Administration (80GSFC18C0011).
The research was partly carried out at the Jet Propulsion Laboratory, California Institute of Technology, under a contract with the National Aeronautics and Space Administration (80NM0018D0004).

The High Performance Computing resources used in this investigation were provided by funding from the JPL Information and Technology Solutions Directorate.  The authors acknowledge the Texas Advanced Computing Center (TACC) at The University of Texas at Austin for providing computational resources that have contributed to the research results reported within this paper. URL: \url{https://www.tacc.utexas.edu}

This work has made use of data from the European Space Agency (ESA) mission Gaia (https://www.cosmos.esa.int/gaia), processed by the Gaia Data Processing and Analysis Consortium (DPAC, https://www.cosmos.esa.int/web/gaia/dpac/consortium). Funding for the DPAC has been provided by national institutions, in particular the institutions participating in the Gaia Multilateral Agreement.

We wish to thank the Vera C. Rubin Observatory Data Management team for their support for the use of the Rubin Butler and pipeline middleware in our work.

\end{acknowledgments}

\facility{SPHEREx}

\software{astrometry.net \citep{langAstrometry2010},
          astropy \citep{astropy2013,astropy2022},  
          kete \citep{kete},
          Photutils \citep{Bradley2024},
          Rubin Observatory Butler and pipeline middleware \citep{Jenness2022ButlerPipeline}
          scamp \citep{Bertin2006},
          Source Extractor \citep{1996A&AS..117..393B},
          SLURM (\url{https://github.com/SchedMD/slurm}),
          Tractor \citep{tractor16}
        }

\bibliography{spherex}{}
\bibliographystyle{aasjournalv7}



\end{document}

%% file: author_list_l1-3_pipeline_paper.tex
\author[0000-0001-9674-1564]{Rachel~Akeson}%
\affiliation{IPAC, California Insitute of Technology, MC 100-22, 1200 E California Blvd Pasadena, CA 91125, USA}%
\email[show]{rla@ipac.caltech.edu}%

\author[0000-0003-1598-6979]{Gregory P. Dubois-Felsmann}%
\affiliation{IPAC, California Insitute of Technology, MC 100-22, 1200 E California Blvd Pasadena, CA 91125, USA}%
\email[show]{gpdf@ipac.caltech.edu}%
\author[0000-0002-4650-8518]{Brendan~P.~Crill}%
\affiliation{Jet Propulsion Laboratory, California Institute of Technology, 4800 Oak Grove Drive, Pasadena, CA 91109, USA}%
\affiliation{Department of Physics, California Institute of Technology, 1200 E. California Boulevard, Pasadena, CA 91125, USA}%
\email{bcrill@jpl.nasa.gov}%

\author[0000-0002-9382-9832]{Andreas~L.~Faisst}%
\affiliation{IPAC, California Insitute of Technology, MC 100-22, 1200 E California Blvd Pasadena, CA 91125, USA}%
\email{afaisst@caltech.edu}%
\author[0000-0002-0665-5759]{Tamim~Fatahi}%
\affiliation{IPAC, California Insitute of Technology, MC 100-22, 1200 E California Blvd Pasadena, CA 91125, USA}%
\email{tfatahi@caltech.edu}%
\author[0000-0001-9925-0146]{Candice~M.~Fazar}%
\affiliation{School of Physics and Astronomy, Rochester Institute of Technology, 1 Lomb Memorial Dr., Rochester, NY 14623, USA}%
\email{cmfsps@rit.edu}%
\author[0009-0003-5316-5562]{Tatiana~Goldina}%
\affiliation{IPAC, California Insitute of Technology, MC 100-22, 1200 E California Blvd Pasadena, CA 91125, USA}%
\email{tatianag@ipac.caltech.edu}%
\author[0000-0001-5382-6138]{Daniel~C.~Masters}%
\affiliation{IPAC, California Insitute of Technology, MC 100-22, 1200 E California Blvd Pasadena, CA 91125, USA}%
\email{dmasters@ipac.caltech.edu}%
\author[0000-0002-5713-3803]{Christina~Nelson}%
\affiliation{IPAC, California Insitute of Technology, MC 100-22, 1200 E California Blvd Pasadena, CA 91125, USA}%
\email{tinan@caltech.edu}%
\author[0000-0002-5158-243X]{Roberta~Paladini}%
\affiliation{IPAC, California Insitute of Technology, MC 100-22, 1200 E California Blvd Pasadena, CA 91125, USA}%
\email{paladini@ipac.caltech.edu}%
\author[0000-0002-7064-5424]{Harry~I.~Teplitz}%
\affiliation{IPAC, California Insitute of Technology, MC 100-22, 1200 E California Blvd Pasadena, CA 91125, USA}%
\email{hit@ipac.caltech.edu}%
\author[0009-0009-4392-3642]{Gabriela~Torrini}%
\affiliation{IPAC, California Insitute of Technology, MC 100-22, 1200 E California Blvd Pasadena, CA 91125, USA}%
\email{gtorrini@ipac.caltech.edu}%
\author[0009-0004-2580-3624]{Phani~Velicheti}%
\affiliation{IPAC, California Insitute of Technology, MC 100-22, 1200 E California Blvd Pasadena, CA 91125, USA}%
\email{phanisrc@caltech.edu}%
\author[0000-0002-3993-0745]{Matthew~L.~N.~Ashby}%
\affiliation{Center for Astrophysics $|$ Harvard \& Smithsonian, Optical and Infrared Astronomy Division, Cambridge, MA 01238, USA}%
\email{mashby@cfa.harvard.edu}%
\author{Dan~Avner}%
\affiliation{IPAC, California Insitute of Technology, MC 100-22, 1200 E California Blvd Pasadena, CA 91125, USA}%
\email{avner@ipac.caltech.edu}%
\author[0000-0002-2618-1124]{Yoonsoo~P.~Bach}%
\affiliation{Korea Astronomy and Space Science Institute (KASI), 776 Daedeok-daero, Yuseong-gu, Daejeon 34055, Republic of Korea}%
\email{ysbach93@gmail.com}%
\author[0000-0002-5710-5212]{James~J.~Bock}%
\affiliation{Department of Physics, California Institute of Technology, 1200 E. California Boulevard, Pasadena, CA 91125, USA}%
\affiliation{Jet Propulsion Laboratory, California Institute of Technology, 4800 Oak Grove Drive, Pasadena, CA 91109, USA}%
\email{jjb@astro.caltech.edu}%
\author[0000-0002-6503-5218]{Sean~Bruton}%
\affiliation{Department of Physics, California Institute of Technology, 1200 E. California Boulevard, Pasadena, CA 91125, USA}%
\email{sbruton@caltech.edu}%
\author[0000-0003-4607-9562]{Sean~A.~Bryan }%
\affiliation{School of Earth and Space Exploration, Arizona State University, 781 Terrace Mall, Tempe, AZ 85287 USA}%
\email{sean.a.bryan@asu.edu}%
\author[0000-0001-5929-4187]{Tzu-Ching~Chang}%
\affiliation{Jet Propulsion Laboratory, California Institute of Technology, 4800 Oak Grove Drive, Pasadena, CA 91109, USA}%
\affiliation{Department of Physics, California Institute of Technology, 1200 E. California Boulevard, Pasadena, CA 91125, USA}%
\email{tzu@caltech.edu}%
\author[0009-0000-3415-2203]{Shuang-Shuang~Chen}%
\affiliation{Department of Physics, California Institute of Technology, 1200 E. California Boulevard, Pasadena, CA 91125, USA}%
\email{schen6@caltech.edu}%
\author[0000-0002-3892-0190]{Asantha~Cooray}%
\affiliation{Department of Physics \& Astronomy, University of California Irvine, Irvine CA 92697, USA}%
\email{acooray@uci.edu}%
\author[0000-0002-7471-719X]{Ari~J.~Cukierman}%
\affiliation{Department of Physics, California Institute of Technology, 1200 E. California Boulevard, Pasadena, CA 91125, USA}%
\email{ajcukier@caltech.edu}%
\author[0000-0001-7432-2932]{Olivier~Dor\'{e}}%
\affiliation{Jet Propulsion Laboratory, California Institute of Technology, 4800 Oak Grove Drive, Pasadena, CA 91109, USA}%
\affiliation{Department of Physics, California Institute of Technology, 1200 E. California Boulevard, Pasadena, CA 91125, USA}%
\email{olivier.dore@caltech.edu }%
\author[0009-0002-0098-6183]{C.~Darren~Dowell}%
\affiliation{Jet Propulsion Laboratory, California Institute of Technology, 4800 Oak Grove Drive, Pasadena, CA 91109, USA}%
\affiliation{Department of Physics, California Institute of Technology, 1200 E. California Boulevard, Pasadena, CA 91125, USA}%
\email{charles.d.dowell@jpl.nasa.gov}%
\author[0000-0002-3745-2882]{Spencer~Everett}%
\affiliation{Department of Physics, California Institute of Technology, 1200 E. California Boulevard, Pasadena, CA 91125, USA}%
\email{severett@caltech.edu}%
\author[0000-0002-9330-8738]{Richard~M.~Feder}%
\affiliation{University of California at Berkeley, Berkeley, CA 94720, USA}%
\email{rmfeder@berkeley.edu}%
\author[0009-0009-1219-5128]{Zhaoyu~Huai}%
\affiliation{Department of Physics, California Institute of Technology, 1200 E. California Boulevard, Pasadena, CA 91125, USA}%
\email{zhuai@caltech.edu}%
\author[0000-0001-5812-1903]{Howard~Hui}%
\affiliation{Department of Physics, California Institute of Technology, 1200 E. California Boulevard, Pasadena, CA 91125, USA}%
\email{hhui@caltech.edu}%
\author[0000-0002-2770-808X]{Woong-Seob~Jeong}%
\affiliation{Korea Astronomy and Space Science Institute (KASI), 776 Daedeok-daero, Yuseong-gu, Daejeon 34055, Republic of Korea}%
\email{jeongws@kasi.re.kr}%
\author[0000-0003-3574-1784]{Young-Soo~Jo}%
\affiliation{Korea Astronomy and Space Science Institute (KASI), 776 Daedeok-daero, Yuseong-gu, Daejeon 34055, Republic of Korea}%
\email{stspeak@kasi.re.kr}%

\author[0009-0003-8869-3651]{Phil~M.~Korngut}%
\affiliation{Department of Physics, California Institute of Technology, 1200 E. California Boulevard, Pasadena, CA 91125, USA}%
\email{pkorngut@caltech.edu}%
\author[0000-0002-8122-3606]{Yuna~G.~Kwon}%
\affiliation{IPAC, California Institute of Technology, 1200 E. California Boulevard, Pasadena, CA 91125, USA}%
\email{ynkwontop@gmail.com}%
\author[0000-0003-1954-5046]{Bomee~Lee}%
\affiliation{Korea Astronomy and Space Science Institute (KASI), 776 Daedeok-daero, Yuseong-gu, Daejeon 34055, Republic of Korea}%
\affiliation{IPAC, California Insitute of Technology, MC 100-22, 1200 E California Blvd Pasadena, CA 91125, USA}%
\email{bomee@kasi.re.kr}%
\author{Gary~J.~Melnick}%
\affiliation{Center for Astrophysics $|$ Harvard \& Smithsonian, Optical and Infrared Astronomy Division, Cambridge, MA 01238, USA}%
\email{gmelnick@cfa.harvard.edu}%
\author[0009-0002-0149-9328]{Giulia~Murgia}%
\affiliation{Department of Physics, California Institute of Technology, 1200 E. California Boulevard, Pasadena, CA 91125, USA}%
\email{gmurgia@caltech.edu}%
\author[0000-0001-9368-3186]{Chi~H.~Nguyen}%
\affiliation{Department of Physics, California Institute of Technology, 1200 E. California Boulevard, Pasadena, CA 91125, USA}%
\email{chnguyen@caltech.edu}%
\author{Milad~Pourrahmani}%
\affiliation{IPAC, California Insitute of Technology, MC 100-22, 1200 E California Blvd Pasadena, CA 91125, USA}%
\email{milad@ipac.caltech.edu}%
\author[0000-0003-4408-0463]{Zafar~Rustamkulov}%
\affiliation{IPAC, California Insitute of Technology, MC 100-22, 1200 E California Blvd Pasadena, CA 91125, USA}%
\email{zafar@caltech.edu}%
\author[0000-0003-1841-2241]{Volker~Tolls}%
\affiliation{Center for Astrophysics $|$ Harvard \& Smithsonian, Optical and Infrared Astronomy Division, Cambridge, MA 01238, USA}%
\email{vtolls@cfa.harvard.edu}%
\author[0000-0002-9554-1082]{Teresa~Symons}%
\affiliation{IPAC, California Insitute of Technology, MC 100-22, 1200 E California Blvd Pasadena, CA 91125, USA}%
\email{tsymons@ipac.caltech.edu}%
\author[0009-0005-3796-2312]{Pao-Yu~Wang}%
\affiliation{School of Earth and Space Exploration, Arizona State University, 781 Terrace Mall, Tempe, AZ 85287 USA}%
\email{pwang55@asu.edu}%
\author[0000-0003-3078-2763]{Yujin~Yang}%
\affiliation{Korea Astronomy and Space Science Institute (KASI), 776 Daedeok-daero, Yuseong-gu, Daejeon 34055, Republic of Korea}%
\email{yyang@kasi.re.kr}%
\author[0000-0001-8253-1451]{Michael~Zemcov}%
\affiliation{School of Physics and Astronomy, Rochester Institute of Technology, 1 Lomb Memorial Dr., Rochester, NY 14623, USA}%
\affiliation{Jet Propulsion Laboratory, California Institute of Technology, 4800 Oak Grove Drive, Pasadena, CA 91109, USA}%
\email{mbzsps@rit.edu}